\documentclass[aps,prl,twocolumn,preprintnumbers,groupedaddress]{revtex4}

\usepackage{amssymb}
\usepackage{epsfig} 
\usepackage{amsmath,color} 
\usepackage{graphicx} 
\usepackage{ulem}

\usepackage{wrapfig}

\newcommand{\ie}{{\it i.e.,\ }}
\newcommand{\eg}{{\it e.g.}}

\newcommand{\beq}{\begin{equation}}
\newcommand{\eeq}{\end{equation}}
\newcommand{\bea}{\begin{eqnarray}}
\newcommand{\eea}{\end{eqnarray}}
\newcommand{\ba}{\begin{array}}
\newcommand{\ea}{\end{array}}
\newcommand{\bi}{\begin{itemize}}
\newcommand{\ei}{\end{itemize}}
\newcommand{\bn}{\begin{enumerate}}
\newcommand{\en}{\end{enumerate}}
\newcommand{\bc}{\begin{center}}
\newcommand{\ec}{\end{center}}
\renewcommand{\l}{\left}
\renewcommand{\r}{\right}

\newcommand{\eq}[1]{Eq.~(\ref{#1})}

\newcommand{\GeV}{\mathinner{\mathrm{GeV}}}

\begin{document}

\preprint{FTUV-15-12-29}
\preprint{IFIC/15-95}

\title{The (relative) size does not matter in inflation}


\author{Gabriela Barenboim}
\email[]{Gabriela.Barenboim@uv.es}
\author{Wan-Il Park}
\email[]{Wanil.Park@uv.es}
\affiliation{
Departament de F\'isica Te\`orica and IFIC, Universitat de Val\`encia-CSIC, E-46100, Burjassot, Spain}


\date{\today}

\begin{abstract}
We show that a tiny correction to the inflaton potential can make critical changes in the inflationary observables for some types of inflation  models.
\end{abstract}


\maketitle


\section{Introduction}

Inflation \cite{Guth:1980zm,Guth:1979bh} is a key concept of modern cosmology, which provides a simple and compelling solution to the main problems of old BigBang cosmology.
Also, the quantum fluctuations of the inflaton (typically a dynamically rolling scalar field controlling the duration of inflation) are regarded as the most plausible seeds of the structures we observe at the present universe \cite{Mukhanov:1981xt,Mukhanov:1982nu}.
Generically, a realization of an observationally consistent inflation requires slow-roll, \ie an inflaton whose effective mass-square parameter is small as compared to the square of the expansion rate during inflation, and the duration of inflation responsible for our visible universe should be around $50-60$ $e$-folds, depending on the thermal history after inflation \cite{KT}.
In addition, thanks to various precise observations, the inflationary observables are being constrained more and more tightly such that many models have been being ruled out or disfavored (see for example Ref.~\cite{Martin:2013tda}).

Inflaton can be a non-trivial trajectory in a multi-dimensional field space, but a majority of models are contained into the single-field scenario.
Conventionally, in a single-field scenario, there is \textit{a} simple form of the effective inflaton potential.
However, it should be noted that, the inflaton couples to other fields, an imprescindible step in order to reheat the universe to recover the standard hot universe necessary for a successful BigBang nucleosynthesis and later cosmology. Such a coupling(s) are indeed effective sub-leading contribution(s) to the inflaton potential.
These contributions might be smaller than the leading potential by several (or many) orders of magnitude, and one may naively expect that such tiny corrections can be ignored.
This may be true in some cases, but may not be always the case.

In this letter, we show that, a tiny correction can make a critical impact on the dynamics of inflaton,  even if it is  several orders of magnitude smaller than the leading order inflaton potential, altering critically the predictions obtained by analyzing only the leading order inflaton potential.

\section{Models}
For a potential $V$ of a single-field slow-roll inflation scenario, the spectral behavior of the density power spectrum originated from the quantum fluctuations of the inflaton is determined by the first two slow-roll parameters defined as
\beq \label{ep-eta-def}
\epsilon \equiv \frac{1}{2} \l| \frac{M_{\rm P} V'}{V} \r|^2, \quad \eta \equiv \frac{M_{\rm P}^2 V''}{V}
\eeq
where `$\prime$' denotes derivative with respect to the inflaton field $\phi$, which is treated as a real scalar field.
They also determine the duration of inflation, i.e, the number of $e$-foldings which is defined as
\beq
N_e \equiv \int_{t_*}^{t_e} H dt \simeq \frac{1}{M_{\rm P}} \int_{\phi_*}^{\phi_e} \frac{d\phi}{-{\rm Sign}(V') \sqrt{2 \epsilon}}
\eeq 
where the subscripts `$_*$' and `$_e$' represent the time or field value at the horizon exit of a given cosmological scale and at the end of inflation, respectively.
In order to be consistent with observations, any inflation model should have $\epsilon$ and $|\eta|$ less than about $\mathcal{O}(10^{-2})$ for the scales probed by CMB observations (e.g., WMAP \cite{Hinshaw:2012aka} or Planck \cite{Ade:2015lrj}) with $e$-foldings upper-bounded as $N_e \lesssim 60$ \cite{KT}.
It is easy to see that, if $|\eta|$ varies slowly around $\mathcal{O}(10^{-2})$ for most of the $e$-foldings of inflation, it is necessary to have $\epsilon \sim \mathcal{O}(10^{-3}-10^{-2})$ in order not to have too many $e$-foldings.
In other words, only if $|\eta|$ varies rapidly, $\epsilon$ can be smaller than $|\eta|$ by several (or many) orders of magnitude.
Although some large field scenarios share such a feature (see for example \cite{Starobinsky:1980te,Bezrukov:2007ep}), this is mostly the case of small-field inflation scenarios where the excursion of the inflaton is limited to be at most Planckian.

When $\epsilon \ll |\eta|$, the spectral index given by $n_s = 1 - 6 \epsilon + 2 \eta$ is determined nearly only by $\eta$. 
However, although it might be tiny, $\epsilon$ affects critically the dynamics of the inflaton via the equation of motion.
In this circumstance, if there is a correction to the inflaton potential, which is tiny in terms of its magnitude but has  sizable derivatives, its impact on the predictions of $V$ could be critical.
  
Keeping this possibility in mind, we assume that the inflaton potential is given by 
\beq
V = V_{\rm B} + V_{\rm M}
\eeq
where $V_{\rm B}$ and $V_{\rm M}$ are the leading order base potential and a sub-leading correction, respectively.
As the examples of $V_{\rm B}$, we consider a Coleman-Weinberg potential \cite{Coleman:1973jx}
\beq \label{V-cw}
V_{\rm cw} = V_0 \l[ 1 + 4 x^4 \l( \ln x - \frac{1}{4} \r) \r]
\eeq
and  a Hilltop potential of the form \cite{Boubekeur:2005zm} 
\beq \label{V-ht}
V_{\rm ht} = V_0 \l( 1 - x^n \r)^2
\eeq
(modulo a completion term) with $x \equiv \phi/\phi_0$ and $\phi_0 \leq M_{\rm P}=2.4 \times 10^{18} \GeV$ for both potentials, and where only $n>2$ is considered. 
For $V_{\rm M}$, we assume 
\beq \label{VM}
V_{\rm M} = \Lambda \l[ \frac{\cos (\nu x)}{1+x} \r]
\eeq
with $\Lambda \lll V_0$ and $\nu > 1$.
The specific form of $V_{\rm M}$ was chosen for a clear and clean illustration of our argument.
Although it might arise from some heavy physics or non-perturbative effect, the origin of $V_{\rm M}$ is out of the scope of this letter. 

As recently studied again in Ref.~\cite{Barenboim:2013wra}, in a small field regime, in a Coleman-Weinberg potential the $e$-foldings associated with the right value of the spectral index are too large to be consistent with observations, and $\epsilon_{\rm cw}(x_*^{\rm cw})$ is smaller than $|\eta_{\rm cw}(x_*^{\rm cw})|$ by many orders of magnitude. 
In the case of the Hilltop potential defined in \eq{V-ht} with $n>2$, one finds
\beq \label{xstar-ht}
x_*^{\rm ht} \simeq \l[ \frac{1}{2n (n-2) N_{e, \rm ht}} \l( \frac{\phi_0}{M_{\rm P}} \r)^2 \r]^{\frac{1}{n-2}} 
\eeq
where
\beq \label{Ne-ns-ht}
N_{e, \rm ht} \simeq \frac{2(n-1)}{(1-n_s)(n-2)}
\eeq
for $x_*^{\rm ht} \ll 1$.
From \eq{Ne-ns-ht}, one finds that Hilltop potential can accomodate $N_e \lesssim 60$ only if $n=4$ with $n_s \lesssim 0.95$ or $n>4$ for a larger $n_s$ \footnote{There would be some changes in this analysis if Hilltop potential is completed in a different way other than what we do as \eq{V-ht} in this work.}. 
Note that for $2<n\lesssim4$, $x_*^{\rm ht}$ is smaller than unity  at least by several orders of magnitude.
Again, this means that $\epsilon_{\rm ht}(x_*^{\rm ht})$ is smaller than $|\eta_{\rm ht}(x_*^{\rm ht})|$ by several orders of magnitude.
In these cases, those tensions (or the plain inconsistency) with observations may be alleviated (or solved)  in the presence of $V_{\rm M}$, since, even if $V_{\rm M} \lll V_{\rm B}$, $V_{\rm M}$ may provide a sizable (or large) contribution to $V'$ (equivalently to $\epsilon$), reducing the number of $e$-foldings so as for those models to be viable.
Concretely, when $\phi_* \lll \phi_0$ which would be the case for our examples of $V_{\rm B}$, if $\nu x_* \ll 1$, one finds 
\bea
\frac{M_{\rm P} V_{\rm M}'}{V} &\simeq& - \frac{\Lambda}{V} \frac{M_{\rm P}}{\phi_0}
\\
\frac{M_{\rm P}^2 V_{\rm M}''}{V} &\simeq& - \frac{\Lambda}{V} \l( \frac{M_{\rm P}}{\phi_0} \r)^2 (\nu^2-2)
\eea
Hence, if 
\beq
\frac{M_{\rm P} V_{\rm B}'}{V} \lesssim - \frac{\Lambda}{V} \frac{M_{\rm P}}{\phi_0} \lesssim \mathcal{O}(10^{-3}-10^{-2})
\eeq
a sizable additional force acts on the inflaton, terminating inflation earlier.
Note that $\nu$ is constrained not to be too large in order to avoid a too large contribution to $\eta$.
Note also that, as $x$ becomes larger, $\epsilon_B$ and $|\eta_B|$ defined in the way of \eq{ep-eta-def} with $V_{\rm B}$ increase rapidly, dominating over the contributions coming from $V_{\rm M}$.
Hence, even if $V_{\rm M}$ is a oscillatory function, the oscillatory behavior would be negligible as inflaton evolves toward the end point of inflation, depending on $\nu$.

\section{Numerical analysis}
In this section, we present the results of our numerical analysis showing the impact of $V_{\rm M}$ in \eq{VM} on the predictions of $V_{\rm B} = V_{\rm cw}$ and $V_{\rm ht}$.

%
\begin{figure*}[t]
\begin{center}
\includegraphics[width=0.32\textwidth]{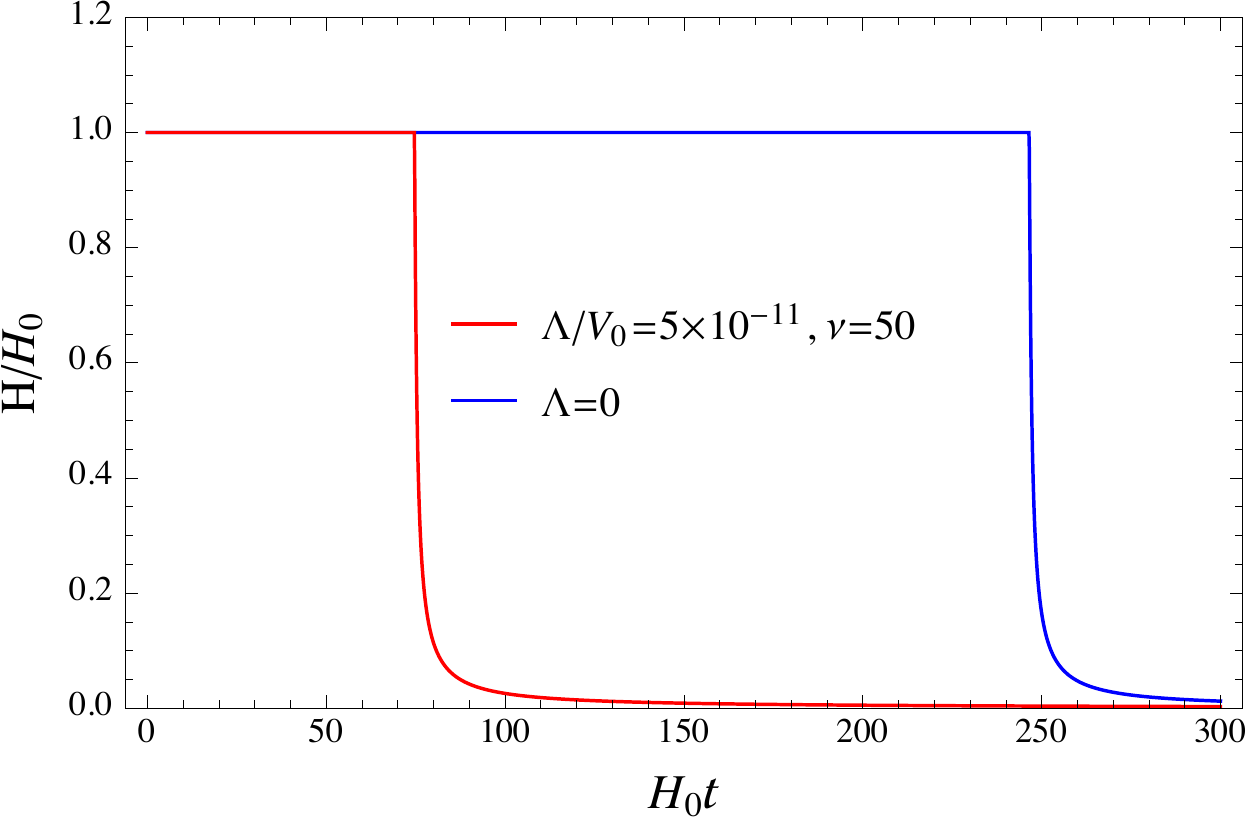}
\includegraphics[width=0.32\textwidth]{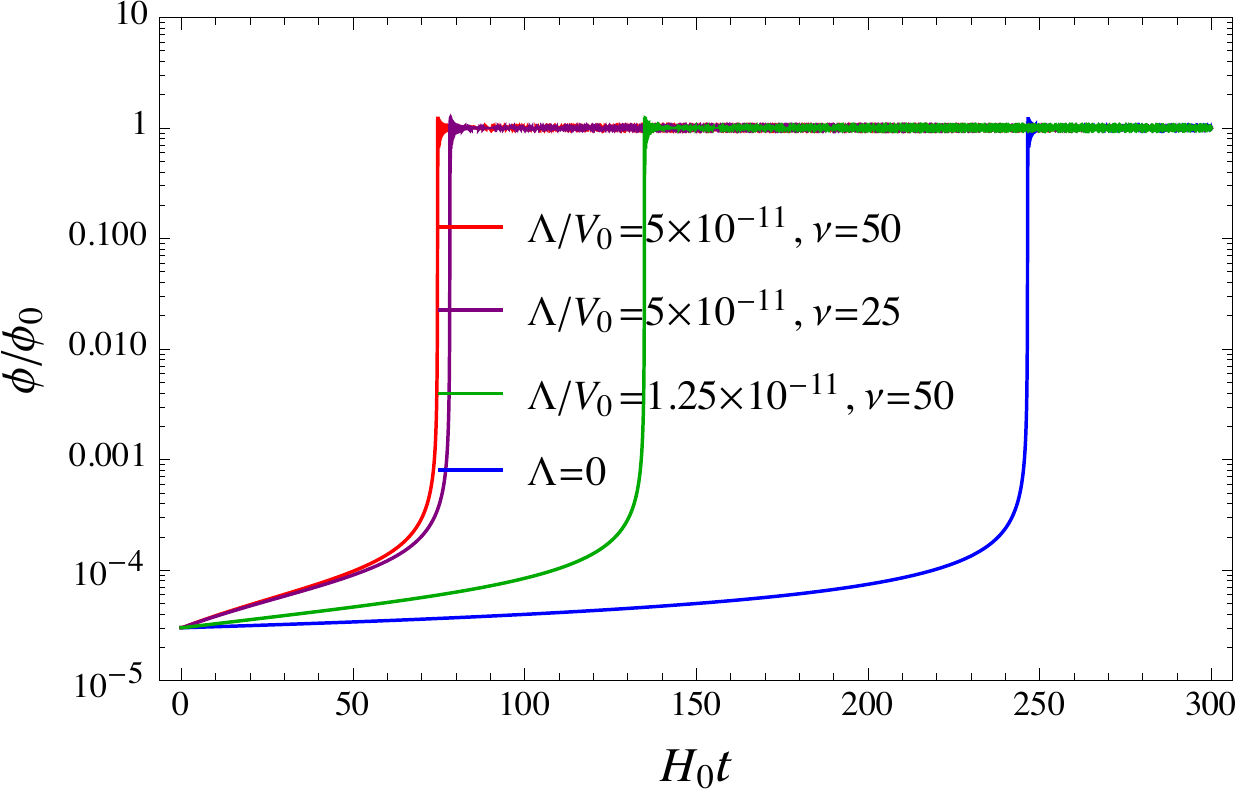}
\includegraphics[width=0.32\textwidth]{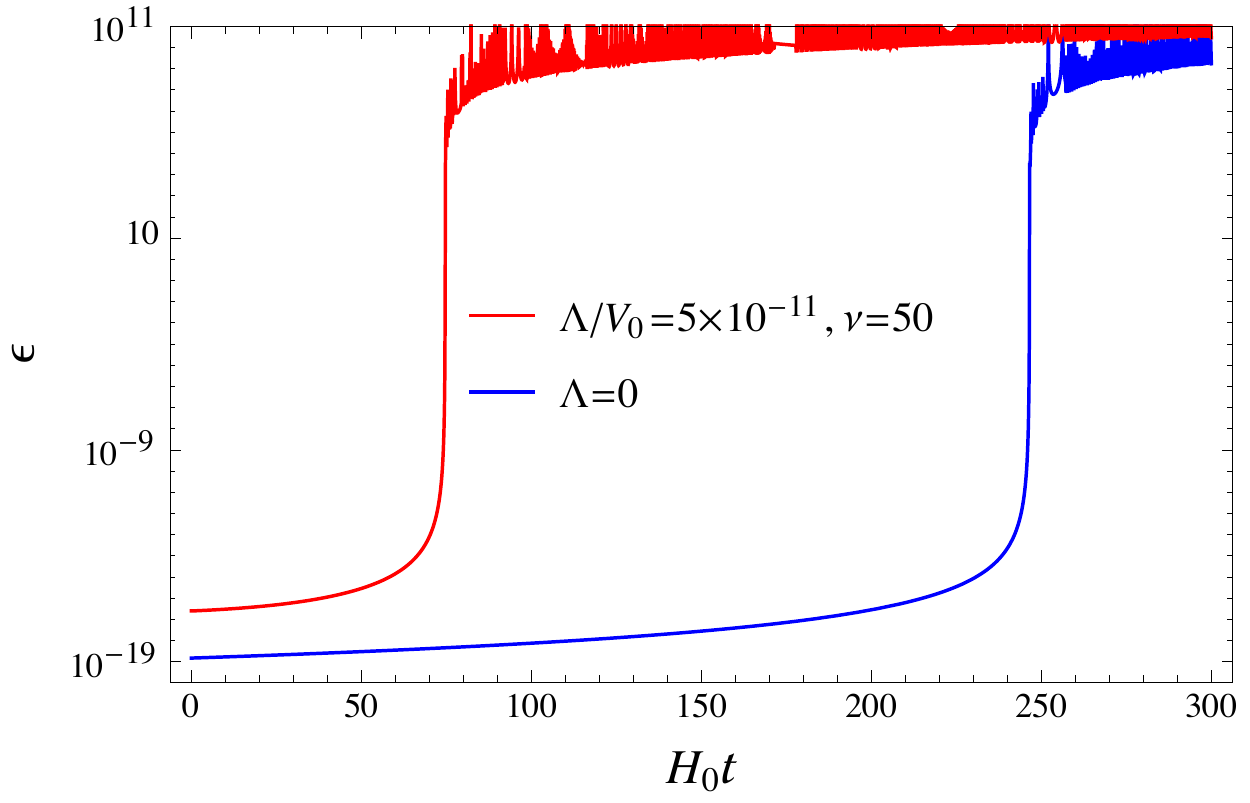}
\includegraphics[width=0.32\textwidth]{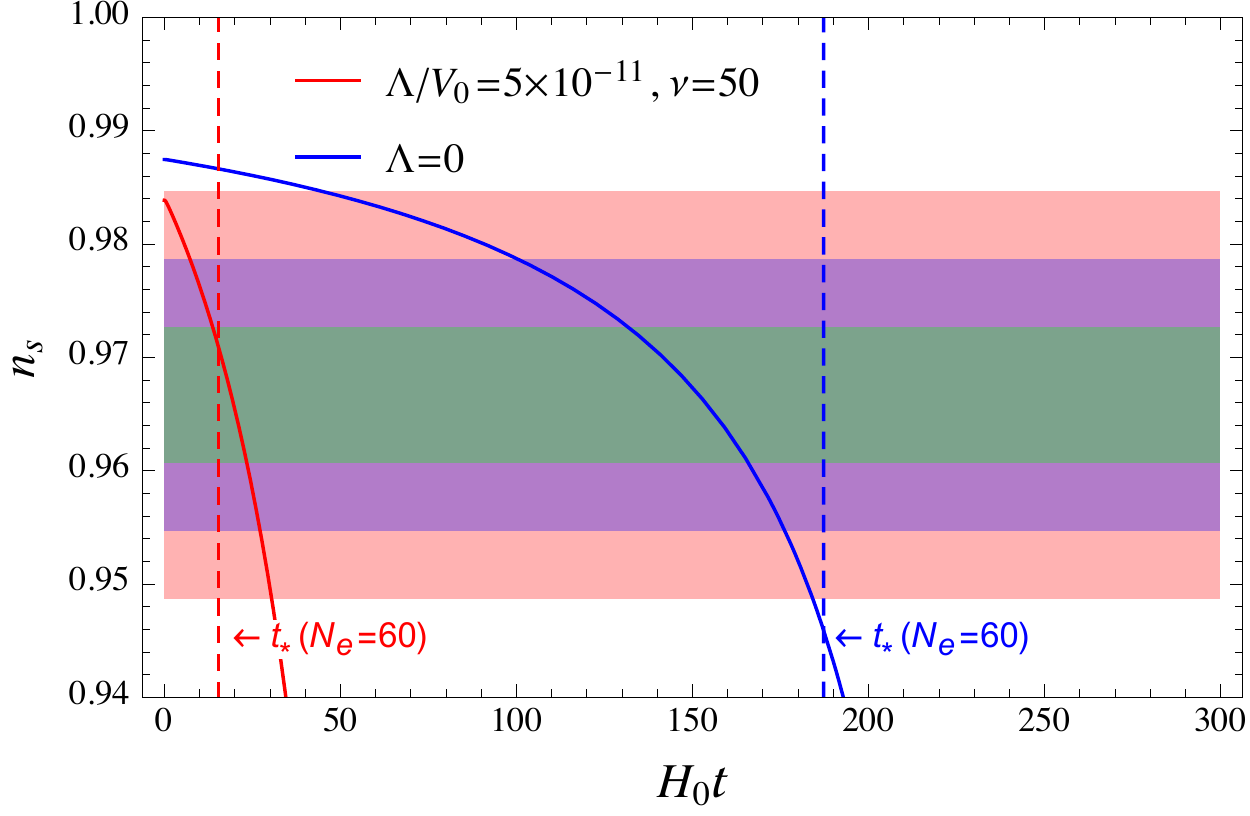}
\includegraphics[width=0.32\textwidth]{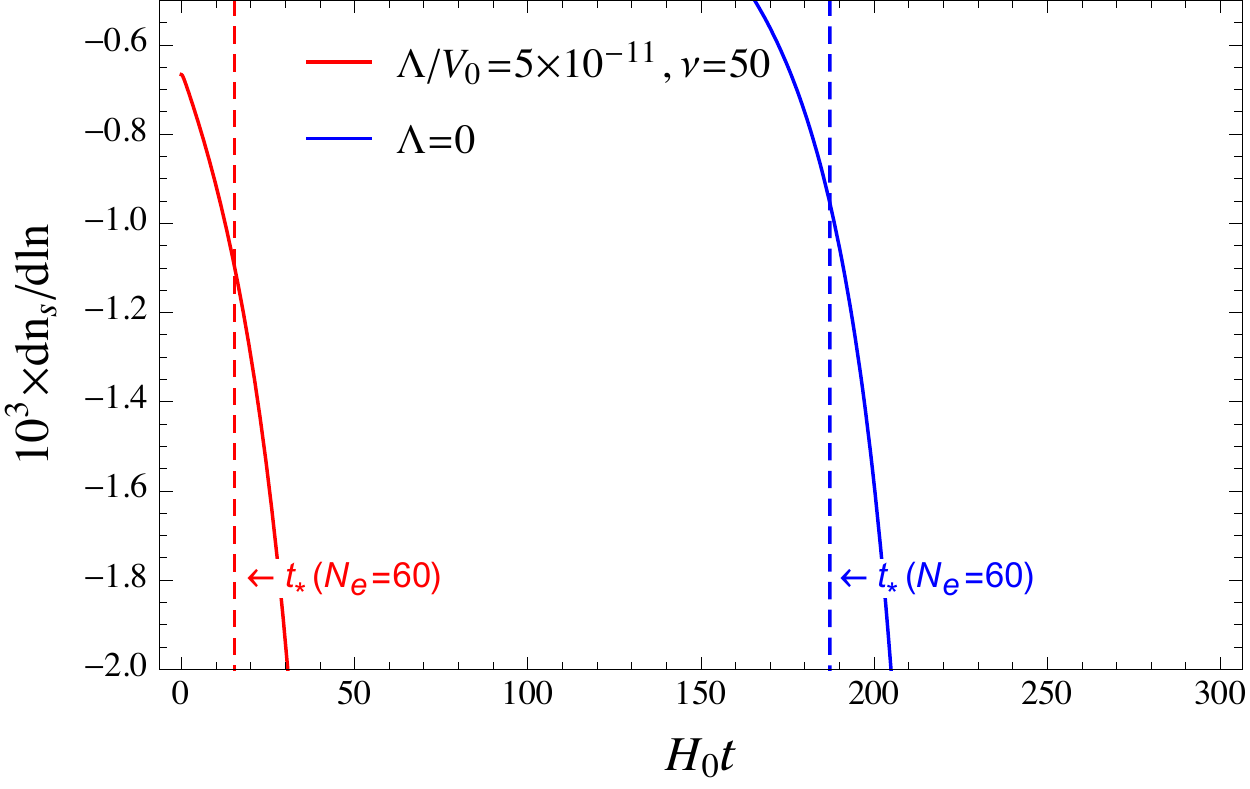}
\includegraphics[width=0.32\textwidth]{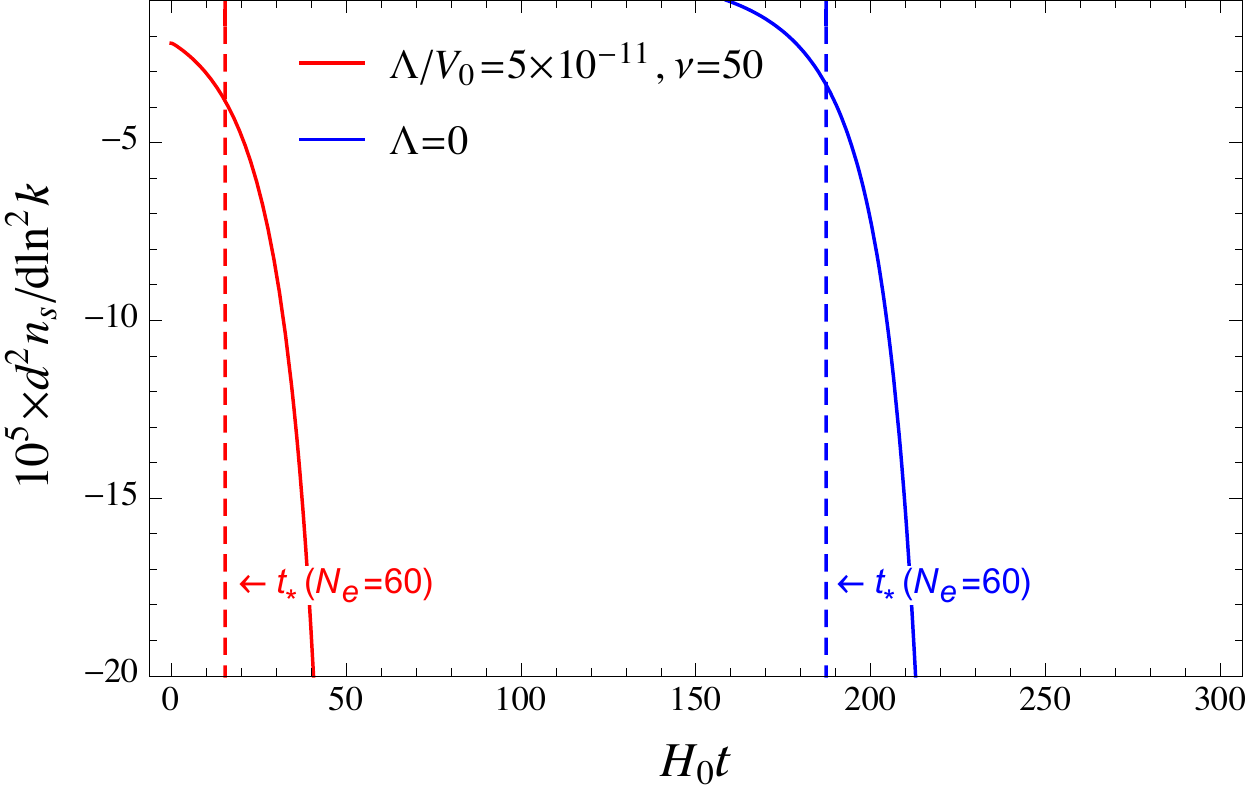}
\caption{The impact of $V_{\rm M}$ for $V_{\rm B}=V_{\rm cw}$ with $\phi_0=M_{\rm GUT}$.
The cosmic time was normalized by $H_0 \equiv V_0^{1/2}/\sqrt{3} M_{\rm P}$.
Colors indicate different combinations of ($\Lambda/V_0, \nu$).
\textit{Top}: Evolution of the Hubble paramere (left), inflaton (middle), and the first slow-roll parameter (right) as functions of time. 
\textit{Bottom}: Solid lines are spectral index (left), the running of the spectral index in unit of $10^{-3}$ (middle), and the running of the running in unit of $10^{-5}$ (right) as functions of time.
Dashed lines correspond to $t_*$ associated with $N_e = 60$ with the same color scheme as solid lines.
Shaded regions in the left one of bottom panels are $1$-, $2$-, and $3$-$\sigma$ uncertainties of $n_s$ as by Planck data \cite{Ade:2015lrj}.
}
\label{fig:cw}
\end{center}
\end{figure*}
%
%
\begin{figure*}[t]
\begin{center}
\includegraphics[width=0.32\textwidth]{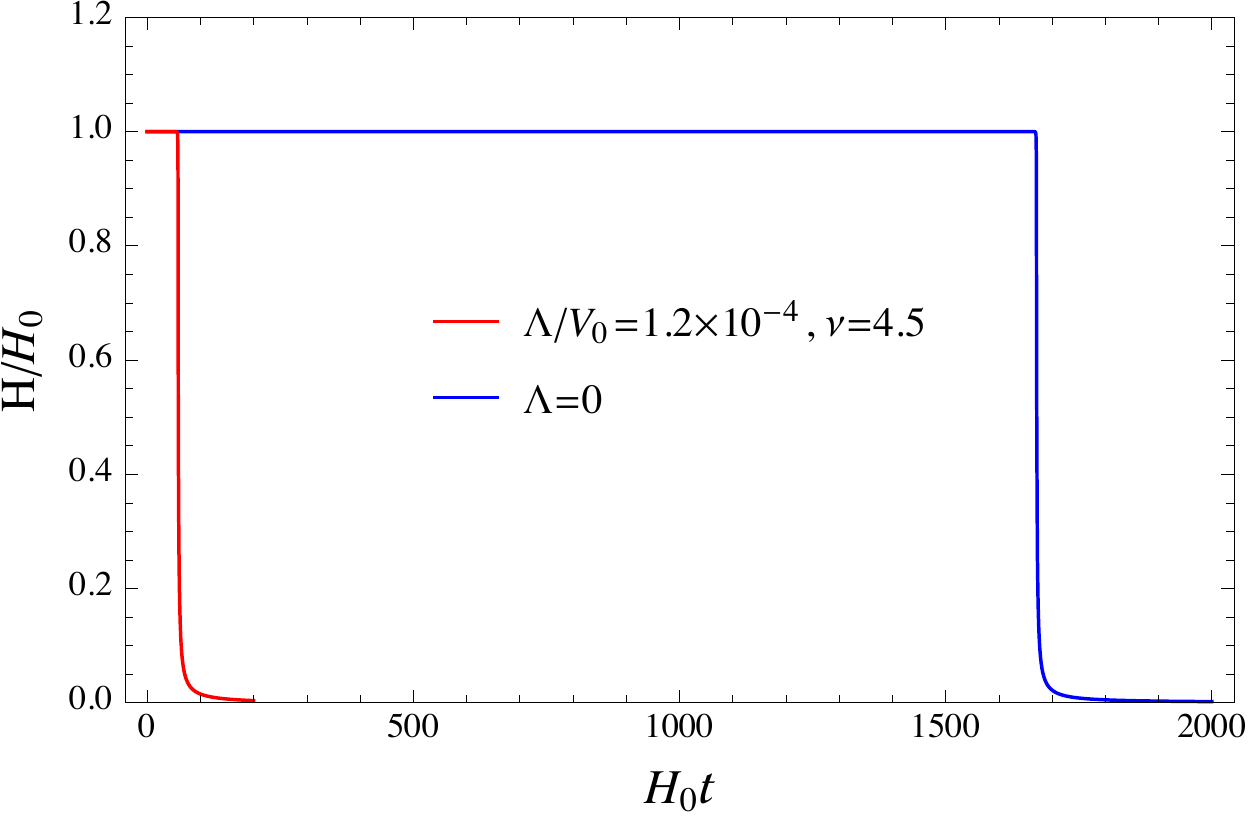}
\includegraphics[width=0.32\textwidth]{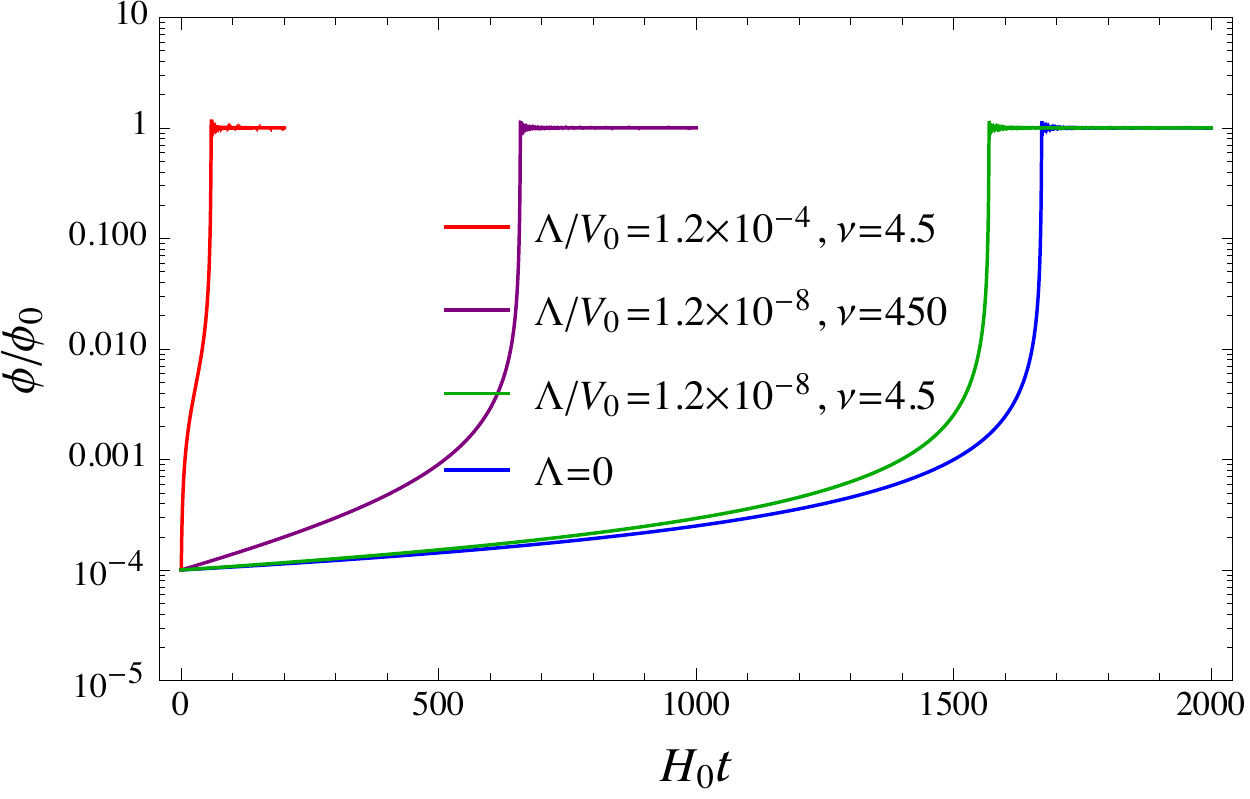}
\includegraphics[width=0.32\textwidth]{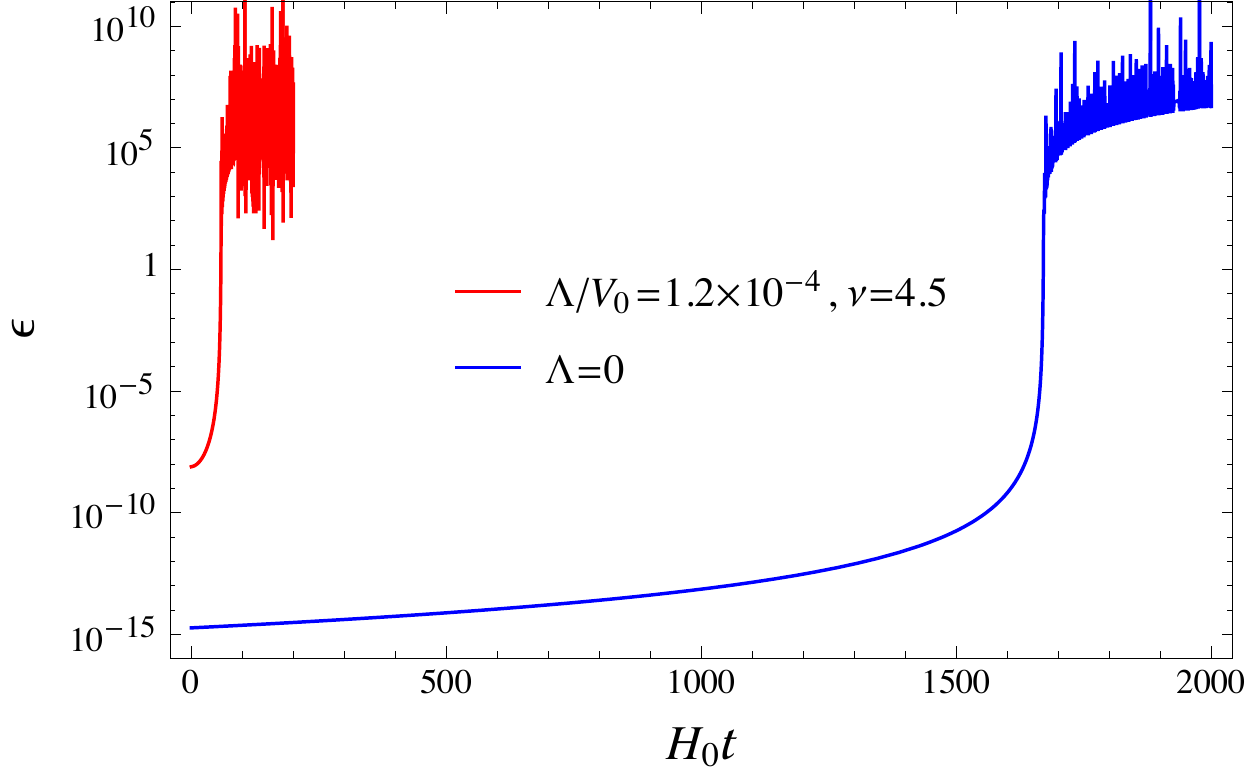}
\includegraphics[width=0.32\textwidth]{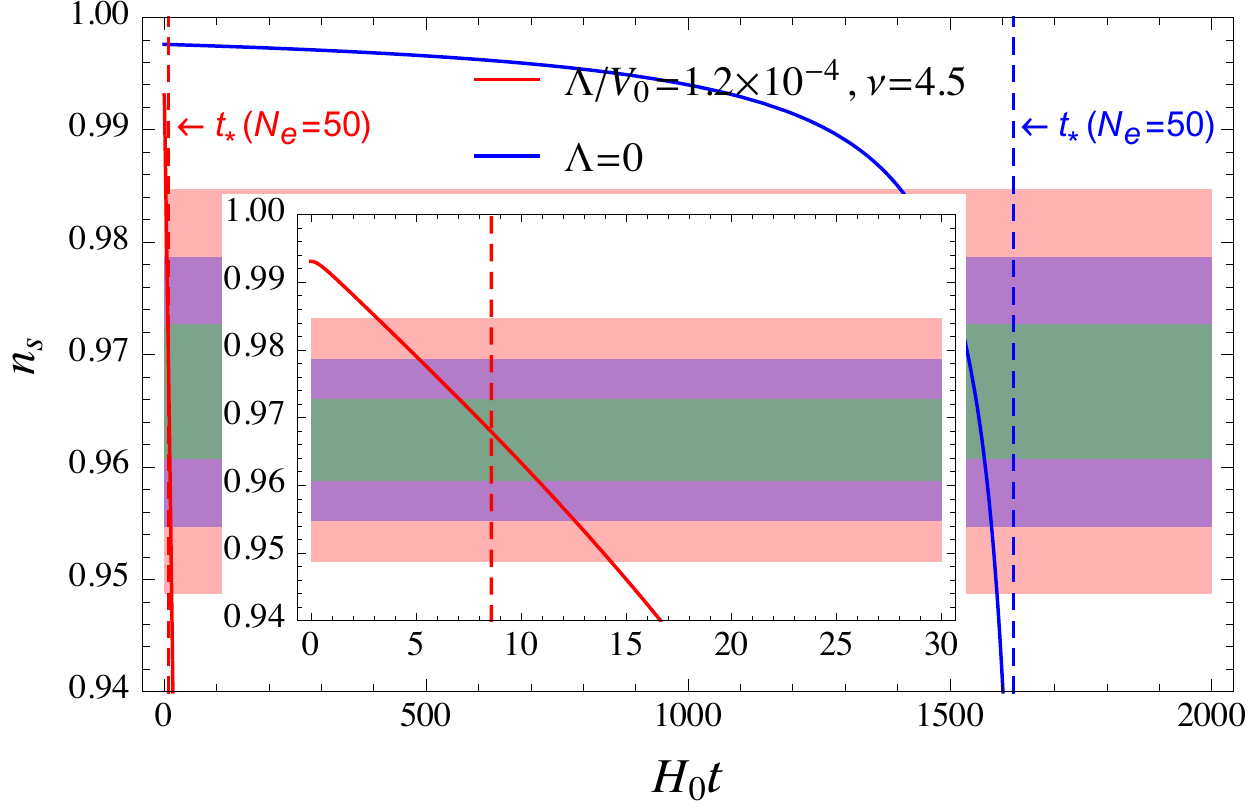}
\includegraphics[width=0.32\textwidth]{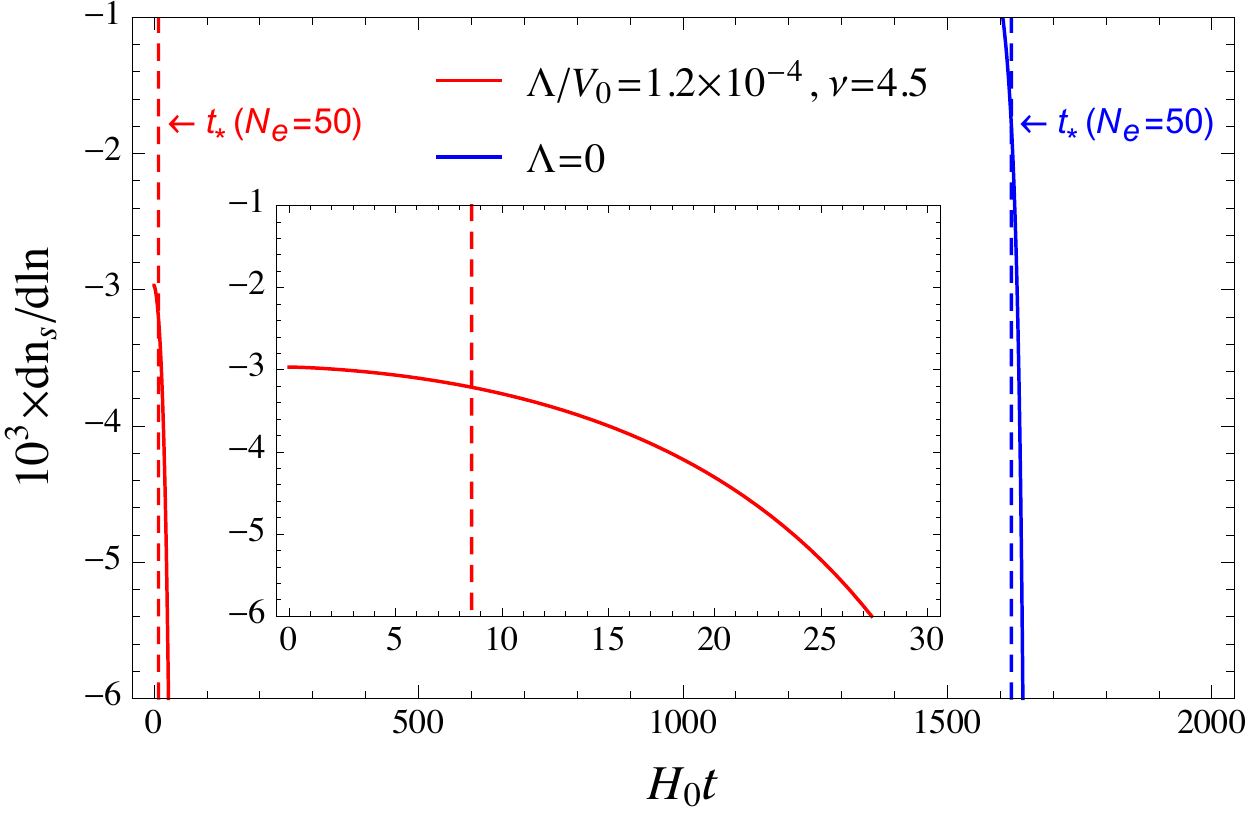}
\includegraphics[width=0.32\textwidth]{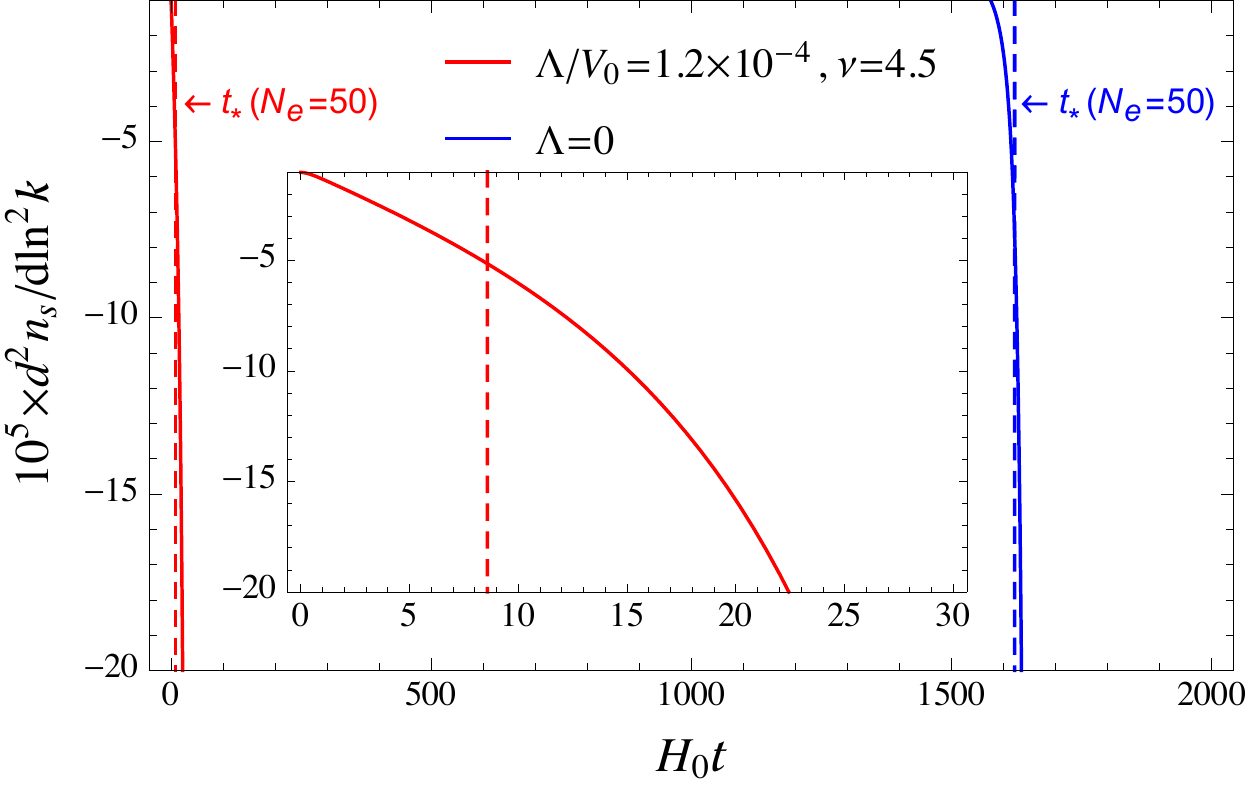}
\caption{The impact of $V_{\rm M}$ for $V_{\rm B}=V_{\rm ht}$ with $n=3$ and $\phi_0=M_{\rm P}$, presented in the same way as Fig.~\ref{fig:cw}.
In the bottom panels, dashed lines correspond to $t_*$ associated with $N_e = 50$ with the same color scheme as solid lines.
The small plot inside the bottom panels are the magnification for a clearer view of the red lines.
}
\label{fig:ht-n3}
\end{center}
\end{figure*}
%
%
In Fig.~\ref{fig:cw}, the case of Coleman-Weinberg base potential is shown for various combinations of ($\Lambda/V_0, \nu$).
As shown in the top left panel of the figure, even if $\Lambda$ is extremely small as compared to $V_0$, it can make a significant change in the dynamics of $\phi$, terminating inflation much earlier.
The evolution of the inflaton is shown in the top middle panel where we plot several combinations of ($\Lambda/V_0,\nu$) represented as colored lines to show the dependence of the inflaton dynamics on the parameters $\Lambda$ and $\nu$.
It can be seen there that the change of $\epsilon$ is more critical than that of $\eta$.
The time dependence of $\epsilon$ is shown in the top right panel of the figure, and our choice of $\Lambda$ increases $\epsilon$ by a factor about $192$ while $\eta$ increases by a factor of $2.2$ at $t_*$ corresponding to $N_e=60$, but $\epsilon(t_*)/\epsilon_{\rm cw}(t_*^{\rm cw}) \simeq 2.68$ and $\eta(t_*)/\eta_{\rm cw}(t_*^{\rm cw}) \simeq 0.54$.
The early termination of inflation requires  the inflaton to be pushed back towards the origin for a given amount of $e$-foldings, resulting in a smaller $\eta$.
Hence, it becomes possible to obtain the observed amount of $n_s$ for the right amount of $e$-foldings, as shown in the bottom left panel of the figure.
As shown in the bottom middle and right panels, there are not significant changes in the running and the running of the running of the spectral index, although the running is pushed slightly to a more negative value.
These weak impacts on the spectral runnings are because the contributions of $V_{\rm M}$ to higher derivatives of $V$ are, at most, comparable to those of the $V_{\rm B}$ we are considering. 
 
In Fig.~\ref{fig:ht-n3}, one can find a similar impact of $V_{\rm M}$ on $V_{\rm B}=V_{\rm ht}$ with $n=3$.
In this case, our choice of $\Lambda$ increased $\epsilon$ by a factor about $5 \times 10^6$ and $\eta$ by a factor about $13$ at $t_*$ associated wtih $N_e=50$, but $\epsilon(t_*)/\epsilon_{\rm ht}(t_*^{\rm ht}) \simeq 3.16$ and $\eta(t_*)/\eta_{\rm ht}(t_*^{\rm ht}) \simeq 0.38$.
As expected, the impact on the running of the spectral index is stronger than the case of $V_{\rm cw}$, but still the change is of a factor less than $2$.
Although we do not present the case here, we found that, $\Lambda/V_0$ in $n=4$ case had to be  slightly increased relative to the case of $n=3$ in order to obtain a similar value of $n_s$ for the same amount of $e$-foldings.
This behavior is due to the fact that, as $n$ increases, $V_{\rm ht}$ becomes flatter towards the origin and allows more $e$-foldings for a given value of $\phi$. 

In all of these cases including the case of Coleman-Weinberg potential, the magnitude of the running is of $\mathcal{O}(1-4) \times 10^{-3}$ for $N_e \sim 40-60$.
Such largish spectral running seems to be a characteristic of small-field inflation models which have rapidly varying potentials. A fact that implies large higher derivatives of the potential, and make them  distinguishable from their large field  competitors\cite{Garcia-Bellido:2014gna,Barenboim:2015lla}

There could be a Hubble-induced mass term in both  $V_{\rm cw}$ and $V_{\rm ht}$ in the sense of supergravity.
However, we found that, as long as the effective mass-square of the quadratic term is less than the square of the expansion rate by about $\mathcal{O}(10^{-2})$ (modulo the negative sign assumed) in order to match observations, it makes only minor changes leading to a slightly smaller $n_s$ but does not modify our findings, since the contribution of the mass term to the slope of potential at the relevant flat region is subleading realative to the contribution of $\Lambda$.
However, note that, when $\Lambda=0$, the term can become the main contribution to the derivative of the inflaton potential and can reduce the $e$-foldings by a large amount, although changes in the observables do not seem significant (or go to wrong direction).    

From these examples, it is clear that, even if it might be extremely small, a correction to the base potential can make critical changes in the inflationary observables predicted from the base potential only.
This can be a generic situation for base potentials which have rapidly varying $\epsilon$ and $\eta$. 
The specific choice $V_{\rm M}$ might be questioned, but its form was designed for a clear illustration of our argument.
The main point we want to deliver is that, whatever the form of $V_{\rm M}$ is, if it can give a sizable contribution at least to the first slow-roll parameter, and can produce significant change in the dynamics of the inflaton such that the predictions of inflationary observables can be critically altered.

\section{Conclusions}

In this letter, we showed that a miserably tiny correction to the inflaton potential can make a significant change in the predictions of inflationary observables.
If the inflaton potential is such that its slope and curvature vary rapidly in a monotonic way, a tiny correction to the potential can give a sizable contribution to the slope and curvature in the very flat region of the potential. Precisely the region that determines the inflationary observables.
In this case, mostly because of the extra force acting on it, the inflaton evolves more rapidly, terminating inflation earlier.    
Hence, the inflaton should be pushed to a flatter region (where the curvature is also smaller) in order to have a given amount of $e$-foldings.
These effects make significant changes in the inflationary observables, especially the spectral index of the density power spectrum. Therefore the status of some models, \eg\ the Coleman Weinberg model, should be revised under this new light.

Typically, only the leading order term of the inflaton potential is considered for the analysis of inflationary observables.
However, our findings imply that in order to predict the observables correctly it is necessary to take into account  the possible corrections to the potential (probably to a level at least several orders of magnitude smaller than the leading order potential), although its impact depends on the specific forms of the base potential and that of the corrections. 

All in all, subleading contributions to the potential do exist and their impact in inflation scenarios (in particular small field inflation models) cannot be underestimated (unless previously studied in depth).

\section{Acknowledgements}
The authors acknowledge support from the MEC and FEDER (EC) Grants SEV-2014-0398 and FPA2014-54459 and the Generalitat Valenciana under grant PROME- TEOII/2013/017.
G.B. acknowledges partial support from the European Union FP7 ITN INVISIBLES (Marie Curie Actions, PITN-GA-2011-289442).

\end{document}